\documentclass[12pt,prd,aps,amssymb,amsmath,tightenlines,showpacs,a4paper]{revtex4}
\def\bea{\begin{eqnarray}}
\def\eea{\end{eqnarray}}
\def\beax{\begin{eqnarray*}}
\def\eeax{\end{eqnarray*}}
\def\half{\frac{1}{2}}

\def\vf{\varphi}
\usepackage{graphics}

\begin{document}
\title{Use of Equivalent Hermitian Hamiltonian for $PT$-Symmetric Sinusoidal Optical Lattices}

\author{H.~F.~Jones\email{h.f.jones@imperial.ac.uk}}

\affiliation{Physics Department, Imperial College, London SW7 2BZ, UK}
\date{\today}

\begin{abstract}
We show how the band structure and beam dynamics of non-Hermitian $PT$-symmetric sinusoidal optical lattices can be approached from the point of view of the equivalent Hermitian problem, obtained by an analytic continuation in the transverse spatial variable $x$. In this latter problem the eigenvalue equation reduces to the Mathieu equation, whose eigenfunctions and properties have been well studied. That being the case, the beam propagation, which parallels the time-development of the wave-function in quantum mechanics, can be calculated using the equivalent of the method of stationary states. We also discuss a model potential that interpolates between a sinusoidal and  periodic square well potential, showing that some of the striking properties of the sinusoidal potential, in particular birefringence, become much less prominent as one goes away from the sinusoidal case.

\end{abstract}

\maketitle

\section{Introduction}

The recent surge of interest in quantum Hamiltonians which are not Hermitian but which nonetheless possess a completely real energy spectrum, due to an unbroken $PT$ symmetry, stems from the pioneering paper of Bender and Boettcher\cite{BB}, in which they showed, by numerical and asymptotic analysis, that the entire class of Hamiltonians
\bea\label{x^N}
H=p^2-(ix)^N
\eea
had that property for $N\ge 2$. Apart from the trivial case $N=2$, the simplest example is for $N=3$, where
$H=p^2 +ix^3$.

Since that initial paper, there has been intensive investigation into the properties of Hamiltonians of this kind, whose progress can be followed in the reviews by Bender\cite{CMBR} and Mostafazadeh\cite{AMR}. We  restrict ourselves here to those features that form the essential background to the present paper.

For a viable framework of quantum mechanics one needs not only a real spectrum but also a probabilistic interpretation. In standard quantum mechanics that is provided by matrix elements of the type $\int \psi^* \hat{A} \chi$, orthogonality of eigenfunctions $\int dx\ \psi^*_1\psi_2=0$,  and the probability density $\psi^*\psi$. In $PT$-symmetric quantum mechanics it is found instead that orthogonality of eigenfunctions takes the nonlocal form
$\int dx\ \psi^*_1(-x)\psi_2(x)=0$, that is, $\int dx\ (\psi_1)_{PT}\psi_2=0$. In the context of quantum mechanics this is a problem because the metric involved in the corresponding normalization integral $\int dx\ (\psi)_{PT}\psi$ is not positive-definite, and we do not in the first instance have a proper probabilistic interpretation.
However, it was subsequently found\cite{BBJC} that another metric could be constructed, with the help of a grading operator $C$, which preserved orthogonality and gave a positive normalization integral $\int dx\ (\psi)_{CPT}\psi$. In contrast to standard quantum mechanics, this metric is not universal, but is dynamically determined by the particular Hamiltonian in question. The calculation of this metric is usually
extremely difficult, and in most cases can only be performed approximately, either through a set of algebraic relations\cite{BBJQ}, or through the use of Moyal brackets\cite{SGF}.

A more general framework, of which $CPT$-symmetry is a special case, was developed by Mostafazadeh\cite{AMh}.
A Hamiltonian $H$ is said to be quasi-Hermitian if it can be related to a Hermitian Hamiltonian $h$
by a similarity transformation:
\bea\label{Htoh}
H=\rho^{-1} h \rho,
\eea
where $\rho$ is a positive-definite Hermitian operator. From this we immediately see that
\bea
H^\dag=\eta H \eta^{-1},
\eea
where $\eta=\rho^2$.
The connection with the $CPT$ formulation is that $\eta$ can be identified as $e^{-Q}$ when $CP$
is written\cite{BBJQ} in the exponential form $CP=e^Q$.
Accordingly $\rho$ can be written as $\rho=e^{-\half Q}$.

The corresponding action of the similarity transformation on states is just
\bea\label{simy}
|\psi\rangle=e^{\half Q} |\vf\rangle,
\eea
where $|\psi\rangle$ is a state of the non-Hermitian system governed by $H$ and $|\vf\rangle$ is the corresponding
state in the Hermitian system governed by $h$. Then in the $H$ system the positive-definite metric given by $\eta$ corresponds to the standard quantum-mechanical metric in the $h$ system.
Thus,
\cite{fn1}
\bea\label{simmy}
\langle \psi_1 |e^{-Q}|\psi_2\rangle
=\langle \psi_1 |e^{-\half Q}(e^{-\half Q}e^{\half Q})e^{-\half Q}|\psi_2\rangle
=\langle \vf_1|\vf_2\rangle.
\eea

A surprising recent development has been the application of these ideas to classical optics\cite{op1}-\cite{op9}.
That such a transfer is possible is due to the fact that under certain approximations the equation of propagation of electromagnetic waves reduces to the paraxial wave equation, which has the same form as the Schr\"odinger equation, but with different roles for the objects appearing there.
The equation takes the form
\bea\label{opteq}
i\frac{\partial\psi}{\partial z}=-\left(\frac{\partial^2}{{\partial x}^2}+V(x)\right)\psi,
\eea
where now $\psi(x,z)$ represents the envelope function of the amplitude of the electric field, where $z$ is a scaled propagation distance, and $V(x)$ is the optical potential, proportional to the variation in the refractive index of the material through which the wave is passing. A complex $V$ corresponds to a complex refractive index, whose imaginary part represents either loss or gain. In principle the loss and gain regions can be carefully configured so that $V$ is $PT$ symmetric, that is $V^*(x)=V(-x)$. There is also a non-linear version of this equation, arising from sufficiently intense beams, where there is an additional term proportional to $|\psi|^2\psi$.

Among many recent papers we may mention linear\cite{op5, op8} and non-linear\cite{op9} two-channel problems, and linear\cite{op3,op6,op7} and non-linear\cite{op2} optical lattices. Ref.~\cite{op3}, where, apart from an overall additive constant, the periodic optical potential was taken to be of the form $V=\half A(\cos{2x}+2i V_0\sin{2x})$, is of particular interest for the present paper, since it is a potential for which the equivalent Hermitian Hamiltonian can readily be constructed.  Figures of the propagation profiles have been given, both below and above the threshold for $PT$-symmetry breaking at $V_0=\half$, showing
unusual features, such as non-reciprocity, power oscillations and bifurcation. In what follows we attempt to cast light on these phenomena from the point of view of the equivalent Hermitian system.

\section{Equivalent Hermitian Hamiltonian}
For the potential used in Ref.~\cite{op3}, the analogue Schr\"odinger equation takes the form
\bea\label{H}
-\psi''-\half A(\cos{2x} + 2i V_0\sin{2x})\psi= -\beta\psi
\eea
for an eigenstate of $H$, with eigenvalue $\beta$ and $z$-dependence $\psi\propto e^{-i\beta z}$.
Below the threshold for $PT$-symmetry breaking, $V_0<\half$, the real and imaginary parts of the potential can be combined into a cosine of complex argument, according to\cite{SG}:
\beax
\cos{2x} + 2i V_0\sin{2x}=\surd{(1-4V_0^2)}\cos(2x-i\theta),
\eeax
where $\theta={\rm arctanh}(2V_0)$.
Thus, the non-Hermitian Hamiltonian
\bea\label{HH}
H = p^2-\half A(\cos{2x} + 2i V_0\sin{2x})
\eea
can be converted into the equivalent Hermitian Hamiltonian
\bea\label{h}
h=p^2-\half A\surd{(1-4V_0^2)}\cos{2x}
\eea
by the complex shift $x\to x+\half i \theta$. This can be implemented by the similarity transformation of Eq.~(\ref{Htoh}), namely
\bea
h=e^{-\half Q}H\ e^{\half Q}
\eea
with $Q=\theta \hat{p}\equiv -i \theta d/dx$, which ensures that the spectra of the two Hamiltonians are identical.

In the symmetry-broken case $V_0>\half$, the corresponding identity is instead
\beax
\cos{2x} + 2i V_0\sin{2x}=i\surd{(4V_0^2-1)}\sin(2x-i\zeta),
\eeax
where $\zeta={\rm arccoth}(2V_0)$. However, in this case we have not gained a great deal from the similarity transformation, since the equivalent Hamiltonian $h$ is itself non-Hermitian.

Finally, at the critical value $V_0=\half$, the equivalent potential vanishes altogether, so that the equivalent theory is simply a free theory, with spectrum $\beta=-k^2$, as has been noted by Longhi\cite{op6}, among others. Part of this spectrum can be observed, in the reduced zone scheme in Fig.~1(b) of Ref.~\cite{op3}. The transformation in this case is a singular one, with $\theta\to\infty$, so the methods used below can not be implemented for this limiting case.
\subsection{Band Structure}
In what follows we shall choose $A=4$, the value taken in Ref.~\cite{op3}. Then the analogue Schr\"odinger equation for $h$ for $V_0<\half$ is the Mathieu equation\cite{AS}:
\bea
\vf''+(a-2q\cos{2x})\vf=0,
\eea
with $q=-\surd{(1-4V_0^2)}$ and $a=-\beta$. In general terms the energy levels can be found by the Floquet method, whereby we take two independent solutions $u_1(x)$ and $u_2(x)$ with the respective initial conditions $u_1(0)=1,\ u'_1(0)=0$ and $u_2(0)=0,\ u'_2(0)=1$ and integrate up to the Brillouin zone boundary at $x=\pi$ to form the discriminant
\bea
D(\beta)= \half(u_1(\pi)+u_2'(\pi)).
\eea
If $|D|\leq 1$, there exists a periodic Bloch-Floquet solution of the form
\bea\label{Floquetsoln}
\vf_k(x)&=&c_k u_1(x)+d_k u_2(x),
\eea
satisfying
\bea
\vf_k(x+\pi)&=&e^{i k \pi}\vf_k(x),
\eea
where $k=(1/\pi)\arccos{D}$.
This procedure gives $k$ as a function of $\beta$, a relation that has to be inverted to give the band structure $\beta=\beta(k)$.
In the standard notation for the Mathieu equation, $k(\beta)$ is called the characteristic exponent.
The values of $\beta$ where $k$ is an integer $r$, i.e at the Brillouin zone boundaries, are termed characteristic values, and are of two types, $a_r$ or $b_r$, depending on whether the Bloch wave-function is even or odd.

In fact in Mathematica these functions have been extended to $k$ non-integral (the functions MathieuA and MathieuB), effectively
mapping out the whole band structure $\beta(k)$ without the need to go through the Floquet procedure explicitly. Using this method we show in Fig.~1 the band structure in both the reduced and extended zone schemes for the potential of Eq.~(\ref{h}), or Eq.~(\ref{H}), for $V_0=0.45$.
\begin{figure}[h]
\resizebox{!}{2.5in}{\includegraphics{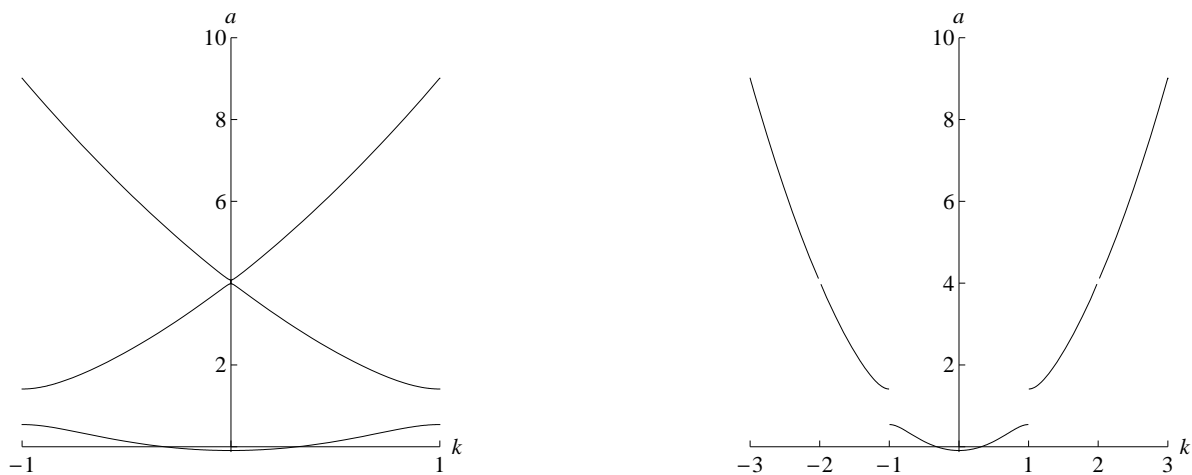}}
\caption{Band structure for $V_0=0.45$ in the reduced and extended zone schemes. In the interests of clarity the gaps at $|k|=2$ have been slightly exaggerated.}
\end{figure}

For $V_0>0.5$ the analogue Schr\"odinger equation for $h$ is
\bea
\vf''+(a+i\surd{(4V_0^2-1)}\sin{2x})\vf=0,
\eea
which, by a shift of $x\to x-\pi/2$, again becomes the Mathieu equation, but with $q$ pure imaginary. The functions $a(k)$ and $b(k)$ are still defined in Mathematica, apart from some minor glitches, and can again be used to map out the band structure.
In this case, the characteristic feature, first observed in \cite{BDM}, is that for real energies, which one is naturally led to consider in solid-state physics, the bands no longer extend to the Brillouin-Zone boundary, but fold back on themselves. In the optical context, however, complex values of $a$ are meaningful, corresponding simply to exponential growth or decay of the beam with $z$. The band structure for $V_0=0.7$, derived using the same method, is shown in Fig.~2.
\begin{figure}[h]
\resizebox{!}{2.5in}{\includegraphics{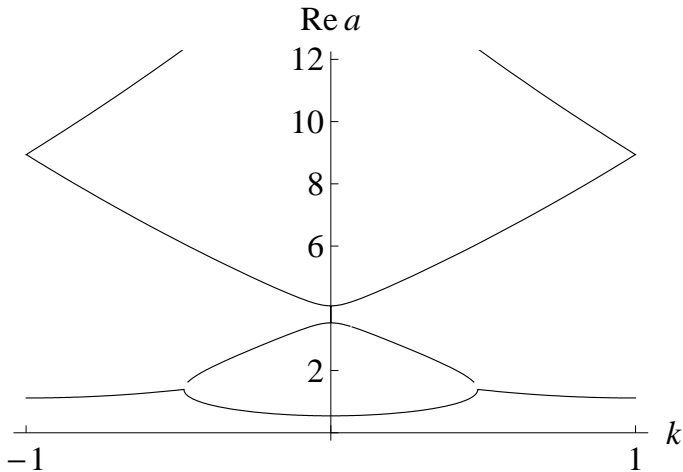}}
\caption{Band structure (real part) for $V_0=0.7$ in the reduced zone scheme.}
\end{figure}

\subsection{Bloch Wave-Functions}
In the present case the Floquet functions $u_1$ and $u_2$ are precisely the even and odd Mathieu functions $ce(a,q,x)$ and $se(a,q,x)$ respectively, up to a normalization factor. For a given value of $k$, we know $a$ and hence can determine the value of the ratio $c_k/d_k$ in the equation (\ref{Floquetsoln}) for the Bloch wave-function. Away from the Brillouin zone boundaries neither $c_k$ nor $d_k$ vanishes, but precisely at those boundaries one or other is zero, making the solution purely symmetric or antisymmetric. Thus, for example, at $k=1$ the wave-function corresponding to the lowest band is symmetric, whereas the wave-function corresponding to the next band is antisymmetric.

The Bloch wave-functions can be individually  normalized in the extended zone scheme according to
\bea
\int_0^\pi |\vf_k(x)|^2 dx=1
\eea
For $k\neq k'$ the orthogonality arises from the different periodicities of  $\vf_k(x)$ and $\vf_{k'}(x)$. If we use periodic boundary conditions in $-N\pi \le x \le N\pi$, so that $k\to k_r=r/N$,
\bea
\int_{-N\pi}^{N\pi}\vf^*_{k_r}(x)\vf_{k_s}(x)\ dx =e^{-\half i \pi\Delta}\left(\frac{\sin{N\pi\Delta}}{\sin{\half\pi\Delta}}\right)
\int_{-\pi}^{\pi}\vf^*_{k_r}(x)\vf_{k_s}(x)\ dx,
\eea
where $\Delta=k_r-k_s=(r-s)/N$. The second factor gives the orthogonality for $r\ne s \mod 2$, while the remaining integral
gives the orthogonality at the BZ boundaries.
\section{Method of Stationary States}
In quantum mechanics a standard method of implementing time development is the method of stationary states. That is, the initial wave-function $\vf(x,t=0)$ is expanded as a superposition of orthonormalized energy eigenstates $\vf_i(x)$:
\bea
\vf(x,t=0)=\sum_i c_i \vf_i(x),
\eea
with
\bea
c_i=\int \vf_i^*(x)\vf(x,t=0) dx,
\eea
and then
\bea
\vf(x,t)=\sum_i c_i \vf_i(x)e^{-E_i t}.
\eea
In the optical problem exactly the same method can be applied, with $z$ taking over the role of $t$, and the eigenstates being the Bloch wave-functions. We first apply this method to the Hermitian problem of Eq.~(\ref{h}) and then show how it can be adapted to give the $z$-development for Eq.~(\ref{H}).
\subsection{Propagation in Hermitian Case}
The initial envelope $\vf(x,z=0)\equiv g(x)$ is to be expanded in terms of the $\vf_{k_r}(x)$, according to
\bea
g(x)=\sum_r c_r \vf_{k_r}(x),
\eea
with the coefficients $c_r$ given by
\bea
c_r=\int_{-N\pi}^{N\pi} \vf^*_{k_r}(x) g(x) dx
\eea
Using the translational property of the Bloch wave-functions we can reduce the integration range to the standard cell $0$ to $\pi$:
\bea
c_r=\int_0^\pi \vf^*_{k_r}(x) G(x) dx\ ,
\eea
where $G(x)=\sum_{q=-N}^{N-1}e^{-i\pi q k_r} g(x + q \pi)$,
and then $\vf(x,z)$ is given by
\bea\label{SSh}
\vf(x,z)=\sum_r c_r \vf_{k_r}(x)e^{-ia(k_r)}.
\eea

For definiteness let us take $g(x)$ to be a broad Gaussian, $g(x)=e^{-(x/w)^2}$,
with $w=6\pi$\ , a function used in Refs.~{\cite{op3},\cite{op6}}. In this case the absolute values of the coefficients
are given in Fig.~3, which reveals that they fall off rapidly with $|k|$, and are essentially negligible for $|k|>3$. They are concentrated around the even integers, reflecting the slowly-varying nature of $g$. Note that by convention the wave-function
is taken to be $ce_k(x)$ at positive integers $k$ and $se_k(x)$ at negative integers.
\begin{figure}[h]
\resizebox{!}{2.5in}{\includegraphics{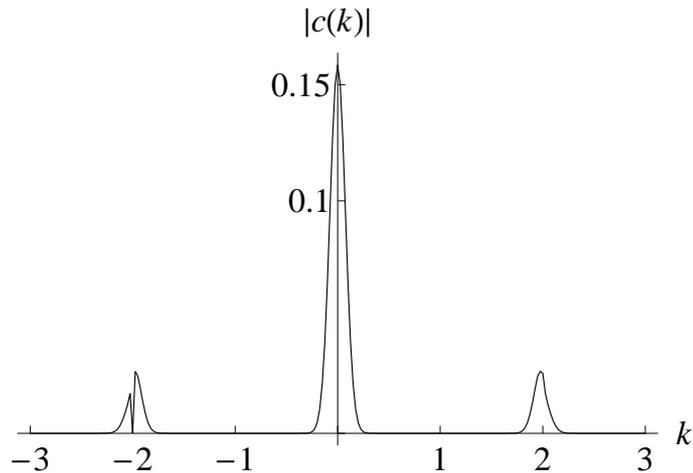}}
\caption{Absolute value of the coefficients $c(k)$ for the Hermitian case.}
\end{figure}

When the sum of Eq.~(\ref{SSh}) is performed, the development of the intensity, shown in Fig.~4, shows no surprises: the beam,
representing a Gaussian wave-front at normal incidence, essentially propagates straight ahead, with a small amount of
lateral spreading.
\begin{figure}[h]
\resizebox{13cm}{!}{\includegraphics{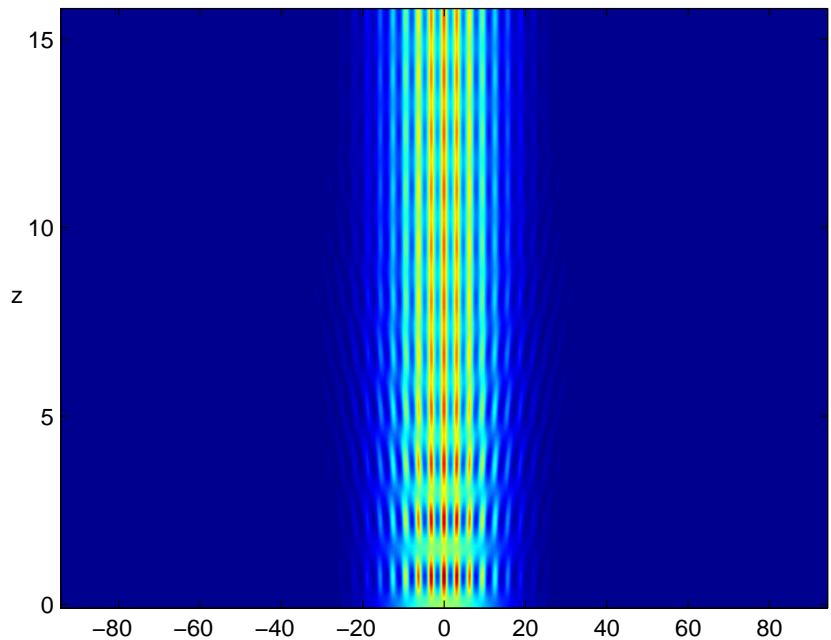}}
\caption{Intensity pattern in the Hermitian case.}
\end{figure}
\subsection{Propagation in non-Hermitian Case}
The situation, however, is very different in the non-Hermitian case.
We now need to fit the initial Gaussian $g(x)$ to the transformed wave-functions $\psi_k(x)$ rather than to the $\vf_k(x)$. That is,
\bea
g(x)=\sum_r d_r \psi_{k_r}(x).
\eea
But from Eq.~(\ref{simy}), which in terms of wave-functions reads $\psi_{k_r}(x)= \vf_{k_r}(x-\half i\theta)$, this can be recast as
\bea
g(x+\half i\theta)=\sum_r d_r \vf_{k_r}(x),
\eea
and then $\psi(x,z)=\vf(x-\half i \theta,z)$.

In fact, for the parameters taken, the coefficients $d_r$ differ very little from the $c_r$. However, because we are plotting $|\psi(x,z)|^2$ rather than $|\vf(x,z)|^2$, the intensity pattern is very different, showing the characteristic birefringence and power oscillations first noted in Ref.~\cite{op3}. Figure 5 is essentially identical to Fig.~2(a) in that paper. It turns out that the asymmetry is primarily due to the contribution of the Bloch functions with $|k|\approx 2$.
\begin{figure}[h]
\resizebox{13cm}{!}{\includegraphics{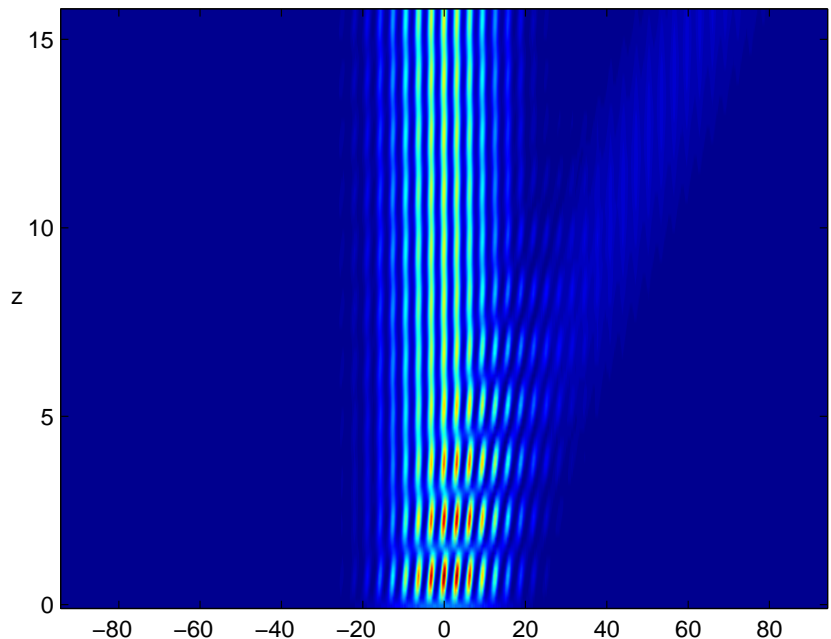}}
\caption{Intensity pattern in the non-Hermitian case.}
\end{figure}
\section{Discussion}

We have shown that it is a simple matter to derive the band structure of non-Hermitian sinusoidal potentials of the type occurring in Eq.~(\ref{H}) using the equivalent Hermitian Hamiltonian of Eq.~(\ref{h}) and the corresponding Mathieu and related functions built in to Mathematica. The dynamics (z-development in optics) can also be implemented using the method of stationary states. Although in practical terms the z-development of the original equation is more efficiently found by direct numerical integration using the split operator method and fast Fourier transform, it is hoped that the stationary state method helps to elucidate the difference between the Hermitian and non-Hermitian situations. In particular it turned out for the parameters we used that the expansion coefficients did not differ significantly in the two cases: the difference between Figs.~4 and 5 was overwhelmingly due to the fact that in the Hermitian case one was plotting $|\vf(x,z)|^2$, whereas in the non-Hermitian case the relevant function was instead the continuation $|\psi(x,z)|^2=|\vf(x-\half\theta,z)|^2$.

One may ask whether the birefringence shown in Fig.~5 is a general feature of $PT$-symmetric potentials, or whether there is something special about the sinusoidal potential. In particular, does the feature persist for a periodic square-well potential (which in practice would be much easier to construct)? The answer seems to be in the negative, and the transition from the sinusoidal case to the square-well case can be neatly studied by using Jacobi $sn$ functions, whose $m$ parameter allows one to interpolate between the two cases. That is, we replace
\bea
V=\half A(\cos{2x}+2iV_0\sin{2x})
\eea
with
\bea\label{Jacobi}
W=\half A\left[{\rm sn}\left(\frac{\pi-4x}{\pi}K(m),m\right)+2i V_0\ {\rm sn}\left(\frac{4 x}{\pi} K(m),m\right)\right],
\eea
rescaling the argument so that the period remains $\pi$. The imaginary part of the potential is shown in Fig.~6 for the two cases $m=0.8$ and $m=0.999$.
\begin{figure}[h]
\resizebox{13cm}{!}{\includegraphics{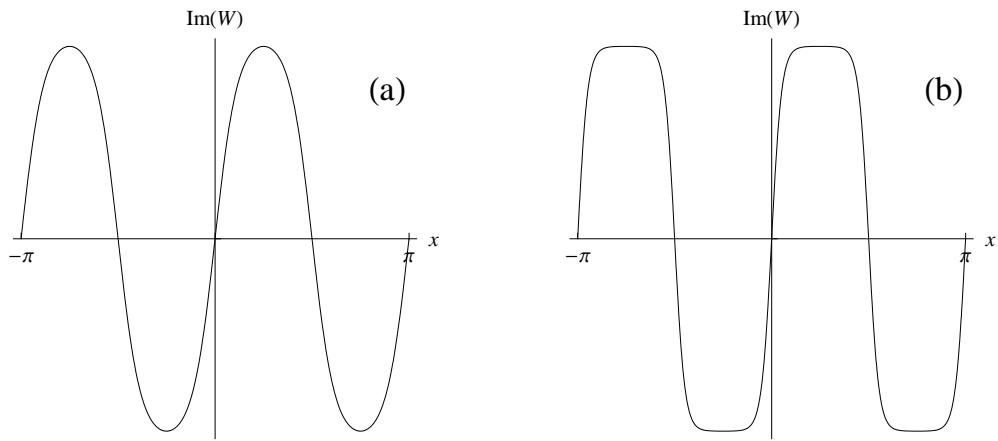}}
\caption{Imaginary part of the potential of Eq.~(\ref{Jacobi}). (a): $m=0.8$, (b): $m=0.999$. }
\end{figure}

For $m=0$ we reproduce Fig.~5, but by the time $m$ reaches 0.8 the birefringence is much less prominent (see Fig.~7, upper panel), and for $m=0.999$, at which point the potential is essentially a periodic square-well, it has more or less disappeared (Fig.~7, lower panel). Since in intermediate cases the wave-functions are not well-known functions, these figures were produced using the original method of direct integration of the differential equation for the $z$-development.

\begin{figure}[h]
\resizebox{12cm}{!}{\includegraphics{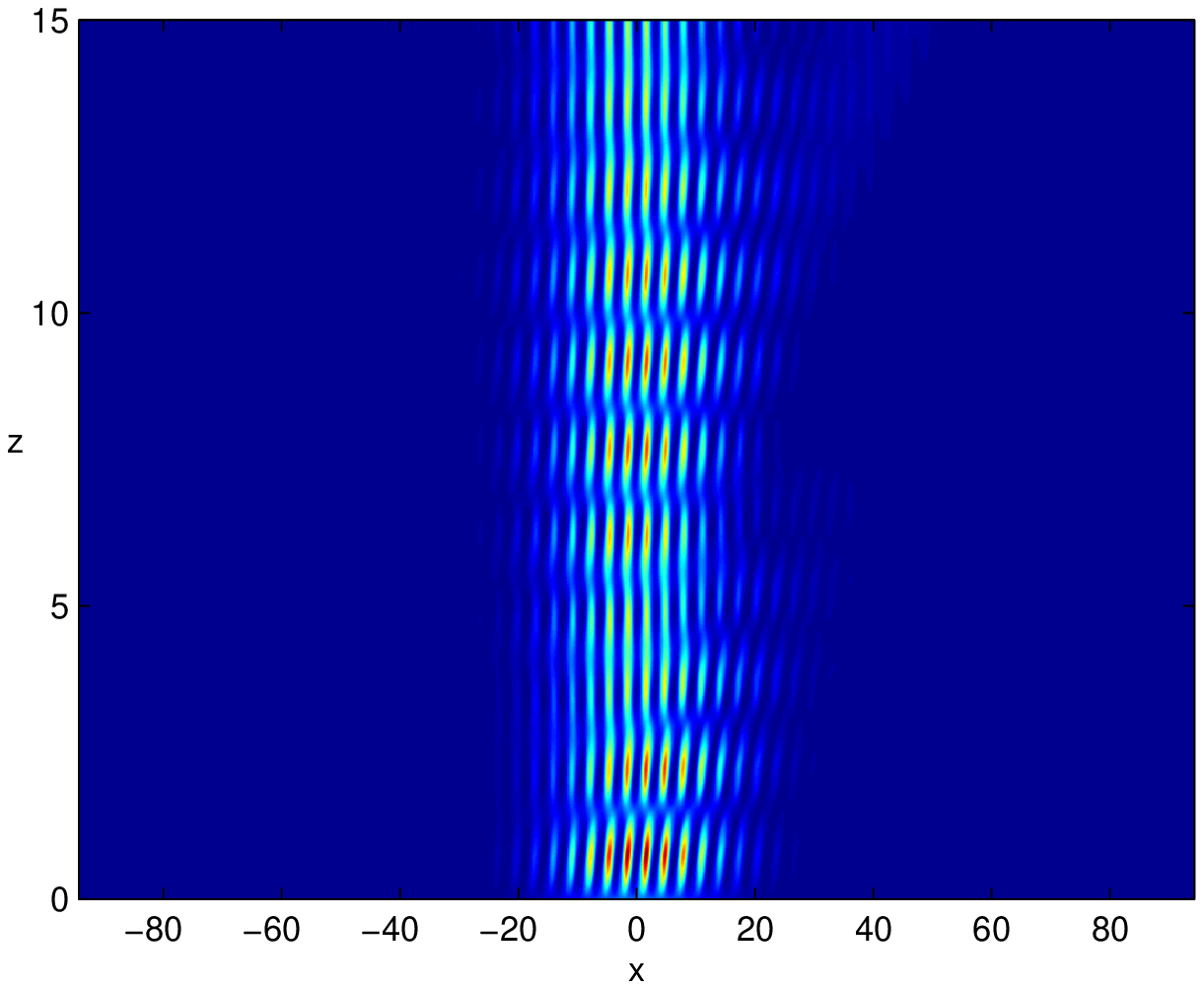}}
\resizebox{12cm}{!}{\includegraphics{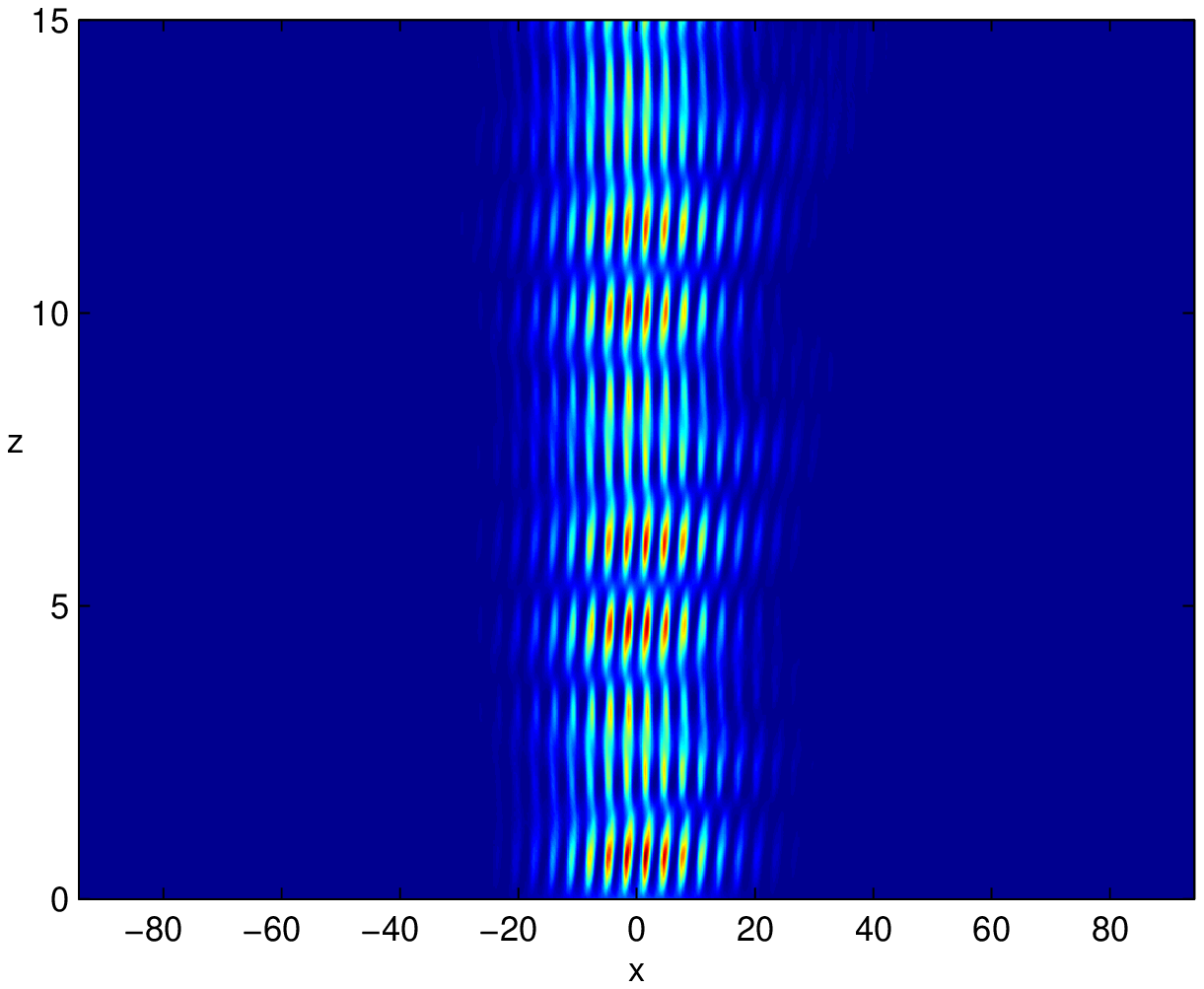}}
\caption{Intensity pattern for $W$ (Eq.~(\ref{Jacobi})). Upper panel: $m=0.8$, lower panel $m=0.999$.}
\end{figure}
A few final remarks about the limiting case $V_0=1/2$: for this particular value the eigenfunctions for the non-Hermitian $V$ are
again known functions, in fact modified Bessel functions $I_k(\sqrt{2}\ e^{ix})$, so that one might hope that the stationary state method
could here be used directly for the $V$ itself. However, as was pointed out by Longhi\cite{op6}, this is not possible because of spectral singularities, or the non-completeness of the Bessel functions at the B-Z boundaries $k=$ integer. This is another manifestation of the singular nature of the similarity transformation between the non-Hermitian and Hermitian problem in this case.

\acknowledgments{I am grateful to Dr.~E.-M.~Graefe and Prof.~C.~M.~Bender for extremely useful conversations and suggestions.}


\begin{thebibliography}{99}
\bibitem{fn1} Note that if $Q$ had been anti-Hermitian, giving a unitary transformation instead
of a similarity transformation, the first factor of $e^{-\half Q}$ would have become $e^{+\half Q}$,
thus cancelling with the second, and leaving no metric factor.
\end{thebibliography}
\end{document}